\documentclass[aps,twocolumn]{revtex4}
\usepackage{graphicx,amsmath,amssymb}

\begin{document}
\title{Properties of entangled photon pairs generated 
by a CW laser with small coherence time: theory and experiment}
\author{Simone Cialdi} \email{simone.cialdi@mi.infn.it}
\affiliation{INFN, Sezione di Milano, I-20133, Italy}
\author{Fabrizio Castelli}\email{fabrizio.castelli@mi.infn.it}
\affiliation{INFN, Sezione di Milano, I-20133, Italy}
\affiliation{Dipartimento di Fisica, Universit\`a di Milano, I-20113, Italy}
\author{Matteo G. A. Paris}\email{matteo.paris@fisica.unimi.it}
\affiliation{Dipartimento di Fisica, Universit\`a di Milano, I-20113, Italy}
\affiliation{CNISM, Udr Milano Universit\`a, I-20113 Milano, Italy}
\affiliation{ISI Foundation, I-10133 Torino, Italy}
\begin{abstract}
The generation of entangled photon pairs by parametric
down--conversion from solid state CW lasers with small
coherence time is theoretically and experimentally
analyzed. We consider a compact and low-cost setup
based on a two-crystal scheme with Type-I phase
matching. We study the effect of the pump coherence
time over the entangled state visibility and over
the violation of Bell's inequality, as a function
of the crystals length. The full density matrix is
reconstructed by quantum tomography. The proposed
theoretical model is verified using a purification
protocol based on a compensation crystal.
\end{abstract}
\maketitle
\section{Introduction}
Generation of entanglement is the key ingredient of quantum
information processing. In optical implementation with discrete
variables the standard source of entangled photon pairs is
parametric down--conversion in nonlinear crystals pumped by
single-mode laser \cite{kwi99}. Recent advances in laser diodes
technology allow the realization of simpler and cheaper apparatuses
for the entanglement generation \cite{deh02, cia}, though the
quality of the resulting photon pairs is degraded by the small
coherence time of the pump laser.
\par
In this paper we address theoretically and experimentally the
generation of entanglement using laser diode pump as well as its
application to visibility and nonlocality tests. We focus on the effects of the
small coherence time and implement a purification protocol based on a
compensation crystal \cite{nambu02} to improve entanglement generation.
We reconstruct the full density matrix by quantum tomography and
analyze in details the properties of the generated state, including
purity and visibility, as a function of the crystals length and the
coherence time of the pump. The topics is relevant for applications
for at least two reasons. On
one hand quantifying the degree of entanglement is of interest in
view of large scale application. On the other hand a detailed
characterization of the generated state allows one to suitably
tailor entanglement distillation protocols.

The paper is structured as follows: In Section \ref{s:exp} we describe the
experimental apparatus used  to generate entanglement, whereas Section
\ref{s:wav} is devoted to illustrate in details the quantum state of the
resulting photon pairs in the ideal case. The effects of small coherence
time are analyzed in Section \ref{s:pol} and the experimental
characterization of the generated states is reported in Section
\ref{s:tom}. Section \ref{s:bel} is devoted to nonlocality test whereas
Section \ref{s:out} closes the paper with some concluding remarks.

\section{The experimental apparatus}
\label{s:exp}
\begin{figure}[!htb]
\centering
\includegraphics[width=0.44\textwidth]{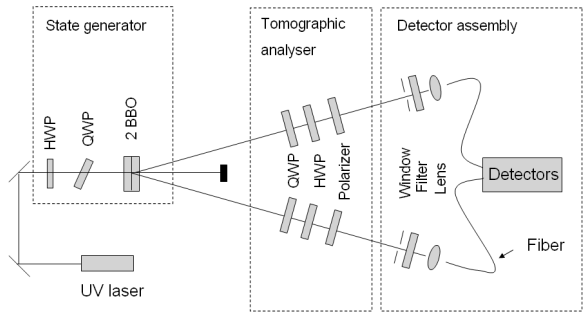}
\caption{Experimental apparatus for generating and analyzing
entangled states.} \label{f:exp}
\end{figure}

A scheme of the experimental apparatus is shown in Fig.~\ref{f:exp}. The
``state generator'' consists of two identical BBO crystals, each cut
for Type-I down-conversion, one half-wave plate (HWP) and one
quarter-wave plate (QWP) as implemented in \cite{gog05}. The crystals are stacked
back-to-back, with their axes oriented at $90^o$ with respect to
each other \cite{kwi99,deh02}. The balancing and the phase of the
entangled states are selected by changing the HWP and QWP orientation.

The crystals are pumped using a 40mW, 405-nm laser diode (Newport
LQC405-40P), with a spectral line that is typically broadened by phonon
collisions. The coherence time of the pump light $\tau_c$, which is a
fundamental parameter for our experiment, results 544 fs and correspond
to a spectral width around 0.3 nm. We obtained this important
information with a standard measurement of the first order correlation
function.
The generated photons are analyzed using adjustable QWP,
HWP and a polarizer \cite{kwi01}. Finally light signals are focused into
multimode fibers which are used to direct the photons to the
detectors. The detectors are home-made single photon counting
modules (SPCM), based on an avalanche photodiode operated in Geiger
mode with active quenching. For the coincidence counting we use a
TAC/SCA.

The nonlinear crystals are properly cut to generate photons into a cone
of half-opening angle $3.0^o$ with respect to pump. The first crystal
converts horizontally polarized pump photons into vertically polarized
($V$) signal and idler photons, while the second crystal converts
vertically polarized pump photons into horizontally polarized ($H$)
signal and idler photons. This configuration introduces a delay time
$\Delta\tau$, depending on the crystal length, between the $V$ and the
$H$ part of the entangled state, as discussed in the following Sections.

\section{The state vector of the generated entangled photons}
\label{s:wav}
The pair of photons generated by SPDC of Type I from a single nonlinear
crystal, having wave vectors $\vec{k}_{s}$ and $\vec{k}_{i}$, are
represented by state vectors $|\vec{k}_{s}\rangle_s$ and
$|\vec{k}_{s}\rangle_i$ for the signal and idler, respectively. The
wavefunction appropriate to the system can be written as a superposition
of these state vectors \cite{hong, joo94, joo96}:
\begin{equation}
 |\Psi\rangle = \int d^3\vec{k}_{s} \, d^3\vec{k}_{i} \,\,
 A(\omega_p - \Omega_p^0) \,
 F(\Delta k_{\perp}) \, f(\Delta k_{\parallel}) \,
 |\vec{k}_{s}\rangle_s \, |\vec{k}_{i}\rangle_i
\end{equation}
where $A(\omega_p - \Omega_p^0)$ is the spectral complex amplitude of the pump laser, which is a function of the pump frequency $\omega_p(k_p)  = \omega_s(k_s) + \omega_i(k_i)$, assuming as usual the validity of the energy conservation in the generation process, and it is centered around the reference frequency $\Omega_p^0$. The factors $F$ and $f$ are mismatch functions depending on the variation of the transverse and longitudinal part of the pump wave vector with respect to the reference of momentum conservation, and are described in detail in the following.

\begin{figure}[h!]
\centering
\includegraphics[width=0.44\textwidth]{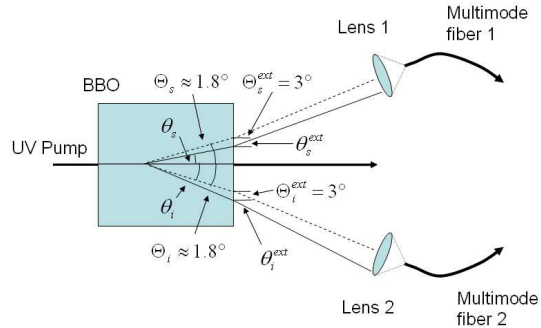}
\caption{Geometry for the generation of photon pairs.}
\label{angoli}
\end{figure}

The function $F(\Delta k_{\perp})$ comes from a spatial integration over all the possible processes of photon generation within the pump transverse profile in the crystal, taking the first order approximation of the nonlinear interaction. For a Gaussian pump profile we obtain again a Gaussian function, with a width varying as the inverse of the beam waist $w$:
\begin{equation}\label{ftrasv}
F(\Delta k_{\perp}) = e^{-w^2 \, \Delta k_{\perp}^2 / 4}
\end{equation}
where, referring to Fig.~\ref{angoli}, one has
\[
    \Delta k_{\perp} = k_s(\omega_s) \sin(\theta_s)
    - k_i(\omega_i) \sin(\theta_i)
\]
with the internal generation angles $\theta_s$ and $\theta_i$ for signal and idler, respectively. In our case the pump beam waist is near 2 mm, therefore we can consider exact transverse momentum conservation to a good approximation. In fact it is easy to verify that with this beam waist we have an angular gaussian width of $0.006^o$ around the reference internal angles $\Theta_s = \Theta_i = 1.8^o$ (derived from external angles $\Theta^{ext}_s = \Theta^{ext}_i = 3.0^o$ using Snell's law), very small with respect to the acceptance angle of $0.074^o$ FWHM of the optical coupling devices. The conservation of the transverse wave vector permits to simplify the geometry of the system, by considering in the following a generation angle, say $\theta_i$, as a function of the other quantities $\omega_s, \omega_i, \theta_s$.

The mismatch function $f(\Delta k_{\parallel})$ has the same meaning of $F$, but derives from an integration along the crystal length $L_C$, and reads:
\begin{equation}\label{ff}
f(\Delta k_{\parallel}) =
    \frac{\sin(\Delta k_{\parallel} L_C /2)}{\Delta k_{\parallel} L_C /2}
\end{equation}
where
\[
    \Delta k_{\parallel} = k_{p}(\omega_p)
    -k_{s}(\omega_s) \cos(\theta_s)
    - k_{i}(\omega_i) \cos\left[\theta_i(\omega_s, \omega_i, \theta_s)\right] \, .
\]
As a matter of fact the pump spectrum width, yet determining the visibility effects, is very small with respect to the spectral width of the down--conversion; this means that $f$ is slightly dependent on $\omega_p  = \omega_s + \omega_i$, as can be verified numerically. We will not consider such a dependence by substituting $\omega_p$ with the reference pump frequency $\Omega^0_p$ as the argument of $f$. This approximation turns out to be very good for crystal lengths below a few mm, but around 3 mm (our maximum crystal length) the conservation of the longitudinal wave vector starts to shrink the down--conversion spectrum. A similar consideration can be done over the dependence of $f$ over the internal angle $\theta_s$; being the experimental configuration highly collinear, the optical couplers are practically insensible to its variation (within the acceptance cone). Therefore we can substitute $\theta_s$ with the fixed reference angle $\Theta_s$, and the mismatch function becomes:
\begin{equation}\label{fo}
    f (\omega_p, \omega_s, \theta_s) \approx f (\Omega^0_p, \omega_s, \Theta_s)
    \equiv f( \omega_s )
\end{equation}
The wavefunction of the photon pair can now be written in the simpler form:
\begin{align}
 |\Psi\rangle = & \int d\omega_p \, d\omega_s \, d\theta_s \,
 A(\omega_p - \Omega^0_p) \,f(\omega_s) \nonumber \\ 
 & \times |\omega_s, \theta_s \rangle_s \,\,
  |\omega_p - \omega_s, \theta_i (\omega_p, \omega_s, \theta_s) \rangle_i
\end{align}

In the first approximation we can solve for the integral over the internal generation angle $\theta_s$ because neither $A$ nor $f$ depend on it, but a more refined reasoning put forward the fact that the conservation of the transverse wave vector introduces a limitation in the effective spectral width of the mismatch function, hence affecting the integration over $\omega_s$. This happens because by varying $\omega_s$ around the down--converted reference $\Omega^0_p/2$, the idler angle $\theta_i$ may go outside from the optical coupler acceptance limit, as verified by means of the experimental data discussed in Appendix~A. This problem does not affect the integral over $\omega_p$ for the smallness of the pump spectral width. To take care of this spectral limitation we introduce a correction factor $R(\Omega^0_p/2, \Delta\omega_s)$ centered around the reference $\Omega^0_p/2$ and having the limited spectral width $\Delta\omega_s$ (see Appendix~A). Defining $\tilde{f}(\omega_s) = f(\omega_s) \cdot R$, we arrive at this wavefunction for the photon pairs:
\begin{equation}\label{wf1}
 |\Psi\rangle = \int d\omega_p \, d\omega_s  \, A(\omega_p - \Omega^0_p) \,
  \tilde{f}(\omega_s) \,\, |\omega_s \rangle_s \,\, |\omega_p - \omega_s \rangle_i
\end{equation}

This expression is used to construct the proper wavefunction (or the proper state vector) for the entangled state generated in our experiment using the pair of oriented crystals \cite{kwi99,deh02}, as described in the previous Section. In particular we consider a suitable superposition of the single crystal wavefunctions of eq.~(\ref{wf1}), introducing the degree of freedom of polarization on state vectors, because the first crystal generates a vertical polarized $(VV)$ and the second crystal generates a horizontal polarized $(HH)$ photon pairs, respectively. Moreover we have a delay time between these pairs, due to the different optical length of the photon trajectories in the inner of crystals. This can be represented in the model by assuming photon generation in the crystals middle \cite{brida1,brida2} (in Appendix~B we show that this is a very good approximation) and introducing propagation factors for the internal state transport.

\begin{figure}[!bht]
\centering
\includegraphics[width=0.44\textwidth]{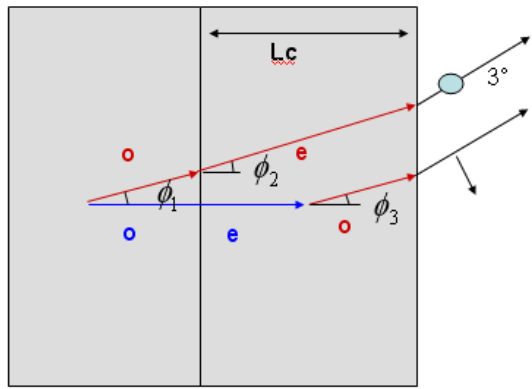}
\caption{Entangled photon generation and propagation inside the
crystals. For clarity, only signal photon trajectories (red lines) are drawn (idler ones are symmetrically upset). The horizontal blue line is the pump ray. $(o)$ and $(e)$ indicates ordinary and extraordinary rays, respectively. $L_c$ is the length of both crystals.}
\label{f:delay}
\end{figure}

In the Fig.~\ref{f:delay} a sketch of the geometry for entangled photon generation is shown, limited for clarity to the signal trajectories. In the first crystal $V$ photons are generated, the state is $|V, \omega_s \rangle_s \,\, |V, \omega_p - \omega_s \rangle_i $, and the complex exponential for the product of the signal and idler propagation factor (as required by the form of eq.~(\ref{wf1})) till to exiting the crystal is given by:
\begin{widetext}
\begin{equation}
  P(V) = \exp \left\{ i L_C \left[ k^o_s(\omega_s)\frac{1}{2 \cos(\phi_1)} +
  k^o_i(\omega_p - \omega_s) \frac{1}{2 \cos(\phi_1)} +
  k^e_s(\omega_s) \frac{1}{\cos(\phi_2)} + k^e_i(\omega_p - \omega_s)
  \frac{1}{\cos(\phi_2)} \right] \right\}
\end{equation}
where the superscripts $(o)$ and $(e)$ on wave vectors indicates ordinary and extraordinary propagation, and the angles $\phi_1 = 1.807, \, \phi_2 = 1.84$ can be found using the laws of wave rays in birefringent crystals \cite{yar} and Snell's law, under the request of an exit angle of $3^o$. For the second crystal, in which $H$ photons are generated and the state is $|H, \omega_s \rangle_s \,\, |H, \omega_p - \omega_s \rangle_i $, the respective propagation factor is:
\begin{equation}
  P(H) = \exp \left\{ i L_C \left[\frac{k^o_p(\omega_p)}{2} +
   \frac{k^e_p(\omega_p)}{2} + k^o_s(\omega_s) \frac{1}{2\cos(\phi_3)} +
   k^o_i(\omega_p - \omega_s) \frac{1}{2\cos(\phi_3)}  \right] \right\}
\end{equation}
where $\phi_3 = 1.806$, and it has been included the propagation of the pump ray from the generation point of the $(VV)$ pair (note that $\phi_3$ is slightly different from $\phi_2$ due the different refraction index for $o$ and $e$ propagation). The entangled state wavefunction is therefore:
\begin{equation}
  | \Psi \rangle  = \, \int \, d\omega_p \, d\omega_s \,
    A(\omega_p - \Omega^0_p) \, \tilde{f}(\omega_s)
    \frac{1}{\sqrt{2}} \left\{ P(H) \, |H, \omega_s \rangle_s |H, \omega_p-\omega_s \rangle_i
   +  P(V) \,
   |V, \omega_s \rangle_s |V, \omega_p-\omega_s \rangle_i \right\}
   \label{propaga}
\end{equation}

This expression can be recast in a more useful form in the following way. Let's write the frequencies as $\omega_p = \Omega_p^0 + \Omega_p$, \, $\omega_s =
\Omega^0 + \Omega$ (with of course $\Omega_p^0 = 2 \Omega^0$), where $\Omega_p$ and $\Omega$ represent the frequency shift with respect to reference for the pump and for the down conversion, respectively. Now, in the propagation factors we introduce a first order approximation for the wave vectors putting:
\[
    k_p(\omega_p) \approx k(\Omega_p^0) + \Omega_p / V_p \,\,, \hspace{1.cm}
    k_s(\omega_s) \approx k(\Omega^0) + \Omega / V \,, \hspace{1.cm}
    k_i(\omega_p - \omega_s) \approx k(\Omega^0)
    + \Omega_p / V - \Omega / V
\]
where $V_p$ and $V$ are the proper group velocities of the pump and of the down converted signal, and these relations must be considered both for the ordinary wave and for the extraordinary wave. With these substitutions, and rewriting for future convenience the quantum states by factorizing the polarization part from the frequency one, the final form of the wavefunction eq.~(\ref{propaga}) read:
\begin{eqnarray}
  | \Psi \rangle   &=& \, \int \, d\Omega_p \, d\Omega \, A(\Omega_p) \,
   \tilde{f}(\Omega^0 + \Omega) \, \frac{1}{\sqrt{2}} \cdot \nonumber \\
   & & \hspace{1.cm} \cdot \left\{ e^{i(\phi_H + \tau_H \Omega_p)} \,
   |H \rangle_s |H \rangle_i |\Omega \rangle_s |\Omega_p-\Omega \rangle_i
   +  e^{i(\phi_V + \tau_V \Omega_p)} \, |V \rangle_s |V \rangle_i
   |\Omega \rangle_s |\Omega_p-\Omega \rangle_i \right\}
   \label{propaga2}
\end{eqnarray}
where the phase terms coming from propagation factors are the sum of a constant phase:
\end{widetext}
\begin{eqnarray*}
  \phi_H &=& \left\{ k^o(\Omega_p^0) + k^e(\Omega_p^0) +
    \frac{2 \, k^o(\Omega^0)}{\cos(\phi_3)} \right\} \frac{L_C}{2} \,\, , \\
    \phi_V  &=& \left\{ \frac{2 \, k^o(\Omega^0)}{\cos(\phi_1)} +
    \frac{4 \, k^e(\Omega^0)}{\cos(\phi_2)} \right\} \frac{L_C}{2}
\end{eqnarray*}
and frequency dependent terms $\tau_H \Omega_p$, $\tau_V \Omega_p$ containing the total propagation time inside the crystals:
\begin{eqnarray*}
   \tau_H  &=& \, \left\{ \frac{1}{V_p^o} + \frac{1}{V_p^e} + \
   \frac{1}{V^o \cos(\phi_3)} \right\} \frac{L_C}{2} \,, \\
  \tau_V  &=& \left\{ \frac{1}{V^o \cos(\phi_1)} + \frac{2}{V^e \cos(\phi_2)}
   \right\} \frac{L_C}{2}
\end{eqnarray*}

It is important to note that these delay factors depend on pump frequency (not on the down converted frequency); this fact can be interpreted saying that the states $(HH)$ and $(VV)$ exiting the crystals are generated from the pump in two different temporal events in the past, depending on the different trajectories across the crystals. For all these four phase factors, their numerical value are determined from the data on refraction indexes and group velocities taken from ref. \cite{snlo}, and listed in the following Table:
\begin{center}
\begin{tabular}{c|c|c|c|c|}
  & \multicolumn{2}{c}{pump} \vline & \multicolumn{2}{c}{signal/idler}
  \vline\\ \cline{2-5}
   & (o) & (e) & (o) & (e) \\ \hline
  $n$ & 1.691719 & 1.659273 & 1.659984 & 1.632171 \\ \hline
  $V$ & c/1.77878 & c/1.73901 & c/1.68376 & c/1.65483 \\ \hline
\end{tabular}
\end{center}

As a final observation, we note that in writing the final expression for the wavefunction eq.~(\ref{propaga2}), it has been discarded the variation of the propagation factors with respect to the propagation angles. Due to small angular acceptance of the detectors, it is possible to verify that, with excellent approximation, this dependence does not introduce any relevant effect.

\section{The polarization density matrix}
\label{s:pol}
For the calculation of the density matrix and the complete characterization of the wavefunction it is important to define at best the statistical properties of the CW pump radiation, because our experimental data depend strongly on its coherence length.
In the temporal domain, this light is characterized by a (real) constant amplitude $A_0$ and a rapidly varying phase with a characteristic time equal to the coherence time of the pump $\tau_c$. Therefore we can write:
\begin{equation}
    \int \, d\omega \, A(\omega) \, e^{i \omega t} \, = \,
    A_0 \, e^{i\delta(t)}
\label{laser}
\end{equation}
where $\delta(t)$ is a proper fluctuating phase. The pump amplitude in the temporal domain can be considered as the Fourier transform of the complex spectral amplitude over a large time interval  $\Delta T$:
\begin{equation}
    A(\omega) \, = \, \frac{1}{2 \pi} \, \int_{\Delta T} \,  d t \, 
    A_0 \, e^{i\delta(t)} \, e^{- i \omega t}
\label{laser2}
\end{equation}

Our experiment mainly concerns the reconstruction of the density matrix of the entangled system on the basis composed by the four signal and idler polarization combinations $HH, HV, VH, VV$. The relative density operator $\rho$, from which we derive the reduced density matrix on this polarization basis, is obtained from the full density operator $\rho_{\mbox{tot}} = | \Psi \rangle \langle \Psi|$ by tracing over frequencies, \emph{e.g.} by integrating over the frequency state matrix elements:

\begin{equation}
\rho = \int d\omega'_p \, d\omega' \, _i\langle \omega'_p - \omega'
| \, _s\langle \omega' | \Psi \rangle \langle \Psi | \omega'
\rangle_s | \omega'_p - \omega' \rangle_i \label{densita1}
\end{equation}
corresponding to the fact that we do not perform frequency measurements.

The form of the wavefunction in eq.~(\ref{propaga2}) implies that only four elements of the $4 \times 4$ reduced density matrix are different from zero. Using the general relation $\langle \omega | \omega^{'} \rangle = \delta (\omega - \omega^{'})$, we straightforwardly obtain for the first diagonal element:
\begin{equation}
\rho_{HH,HH} = \frac{1}{2} \int d\omega \, | f(\omega) |^2 \, \int
d\omega_p \, | A(\omega_p) |^2 \, = \, \frac{1}{2} \, \epsilon \,
A_0^2 \, \frac{\Delta T}{2\pi}
   \label{rhoHHHH}
\end{equation}
where we put $\epsilon = \int |f(\omega)|^2$ and $\int d\omega_p \,
|A(\omega_p)|^2 = A_0^2 \, \Delta T / 2\pi$ from eq.~(\ref{laser2}). With similar calculation the other nonzero diagonal element results $\rho_{VV,VV} = \rho_{HH,HH}$, as expected by symmetry arguments.

For the two off diagonal elements one has $\rho_{HH,VV} = \rho^*_{VV,HH}$, and in particular:
\begin{widetext}
\begin{eqnarray}
  \rho_{HH,VV}  &=& \frac{1}{2} \, \int d\omega | f(\omega) |^2 \, \int d\omega_p 
  \, | A(\omega_p) |^2 \, e^{-i(\phi_H - \phi_V)} \, e^{-i \omega_p(\tau_H - \tau_V)}
    \nonumber \\
   &=& \frac{1}{2} \, \epsilon  \, e^{-i\phi} \int d\omega_p \, |A(\omega_p)|^2 
   \, e^{-i\omega_p (\tau_H - \tau_V)}
\end{eqnarray}
where we put $\phi = \phi_H - \phi_V$. 
With the Wiener-Khinchine theorem this frequency integral can be
recast as a two time correlation function over the interval $\Delta
T$, which can be taken very large with respect to the coherence time of the pump, and
smaller than the detector response time:

\begin{equation}
\int d\omega_p \, | A(\omega_p) |^2 \, e^{-i \omega_p (\tau_H - \tau_V)}
\, = \, A_0^2 \frac{\Delta T}{2\pi} \left( \frac{1}{\Delta T} \int
_{\Delta T} dt \, e^{-i \delta(t) + i \delta \left( t -(\tau_H - \tau_V)\right)}
\right) = A_0^2 \frac{\Delta T}{2\pi} \, e^{- \Delta\tau / \tau_c}
\end{equation}
\end{widetext}
where $\Delta \tau = |\tau_H - \tau_V|$, and the result is taken
from Ref. \cite{blu04}. If $\Delta \tau \gg \tau_c$ we have an
incoherent superposition of random phases and the average of the
complex exponentials tends to zero, otherwise we have a coherent sum, 
and the integral tends to one.

Finally, setting the state generator QWP in order to have $\phi = 0$
(see Ref. \cite{kwi99}) and putting for simplicity $p = e^{-\Delta
\tau /\tau_c}$, the reduced density matrix is:

\begin{equation}\label{matrix2}
\begin{array}{cc}
   & HH \hspace{0.7cm} HV \hspace{0.7cm} VH \hspace{0.7cm} VV \\
   \begin{array}{c}
    HH \\
    HV \\
    VH \\
    VV \\
  \end{array} &
  \left(%
\begin{array}{cccc}
 \frac{1}{2} &  \hspace{0.9cm} 0 \hspace{0.7cm} &
    \hspace{0.7cm} 0 \hspace{0.9cm} & \frac{1}{2} \, p \\
 0 & 0 & 0 & 0 \\
 0 & 0 & 0 & 0 \\
 \frac{1}{2} \, p & 0  &  0 & \frac{1}{2}\\
\end{array}%
\right)
\end{array}
\end{equation}
which can also be conveniently derived from a sum of two distinct
density matrix, one of a pure entangled state and the other of a
statistical mixture: $\rho = p \: \rho_e + (1-p) \rho_m$ where $\rho_e =
| \Psi_e \rangle \langle \Psi_e |$ with $| \Psi_e \rangle =
\frac{1}{\sqrt{2}}(|HH\rangle + |VV\rangle)$ and $\rho_m = \frac{1}{2} |
HH \rangle \langle HH | + \frac{1}{2} | VV \rangle \langle VV |$. This
is the more suitable form used for a comparison between theory and
experimental data.
\section{Experimental tomographic reconstruction of the density matrix and correlation visibility}
\label{s:tom}
In order to fully characterize the generated states at the  quantum
level we employ quantum tomography of their density matrices
\cite{par04}. The experimental procedure goes as follows: upon measuring
a set of independent two-qubit projectors $P_\mu= |\psi_\mu\rangle
\langle\psi_\mu |$ $(\mu=1,...,16)$ corresponding
to different combinations of polarizers and phase-shifters, the
density matrix may be reconstructed as $\varrho=\sum_\mu p_\mu\,
\Gamma_\mu$ where $p_\mu = \hbox{Tr}[\varrho \, P_\mu]$ are the
probabilities of getting a count when measuring $P_\mu$ and
$\Gamma_\mu$ the corresponding dual basis, {\em i.e.} the set of
operators satisfying $\hbox{Tr}[P_\mu\,\Gamma_\nu] =
\delta_{\mu\nu}$ \cite{dar01}.
Of course in the experimental reconstruction the probabilities
$p_\mu$ are substituted by their
experimental samples {\em i.e.} the frequencies of counts obtained
when measuring $P_\mu$. In order to minimize the effects of
fluctuations and avoid non physical results we use
maximum-likelihood reconstruction of two-qubit states
\cite{kwi01,ban00}. At first we write
the density matrix in the form
\begin{equation}
\label{Eq:rhoTT}
\hat{\varrho} \, = \, \hat{T}^{\dagger} \, \hat{T}\;,
\end{equation}
which automatically guarantees that $\hat{\varrho}$ is positive
and Hermitian. The remaining condition of unit trace
$\hbox{Tr}\hat{\varrho} = 1$ will be taken into account using the
method of Lagrange multipliers. In order to achieve the minimal
parametrization, we assume that $\hat{T}$ is a complex lower
triangular matrix, with real elements on the diagonal. This form of
$\hat{T}$ is motivated by the Cholesky decomposition known in
numerical analysis \cite{Cholesky} for arbitrary non negative
Hermitian matrix.  For an $M$-dimensional Hilbert space, the number of
real parameters in the matrix $\hat{T}$ is $M+2M(M-1)/2=M^2$, which
equals the number of independent real parameters for a Hermitian
matrix. This confirms that our parametrization is minimal, up to the
unit trace condition.

In numerical calculations, it is convenient to replace the likelihood
functional by its natural logarithm, which of course does not change the
location of the maximum. Thus the function subjected to numerical
maximization is given by
\begin{equation}
L(\hat{T}) = \sum_{k=1}^N \ln \hbox{Tr}(\hat{T}^\dagger
\hat{T} P_{\mu_k}) - \lambda \hbox{Tr}(\hat{T}^\dagger
\hat{T})\;,
\label{loglik}
\end{equation}
where $\lambda$ is a Lagrange multiplier accounting for normalization
of $\hat \varrho$ that equals the total number of measurements
$N$. This may be easily proved upon writing $\hat{\varrho}$ in
terms of its eigenvectors
$| \phi_\mu
\rangle $ as $
\hat{\varrho} = \sum_{\mu } y_{\mu}^{2} | \phi_\mu
\rangle \langle \phi_\mu |
$, with real $y_{\mu}$, the maximum likelihood condition
$\partial L/\partial y_{\nu} = 0$ reads
\begin{equation}
\lambda y_{\nu} = \sum_{k=1}^{N} \frac{y_\nu \langle \phi_\nu
| P_{\mu_k} | \phi_\nu \rangle}
{\hbox{Tr}(\hat\varrho P_{\mu_k}}\;,
\end{equation}
which, after multiplication by $y_{\nu}$ and summation over $\nu $,
yields $\lambda = N$.

The above formulation of the maximization problem allows one to apply
standard numerical procedures for searching the maximum over the $M^2$
real parameters of the matrix $\hat{T}$. The examples presented below
use the downhill simplex or the simulated annealing methods
\cite{Ameba}.  Results of the reconstruction are reported for crystals
with three different thicknesses, precisely $0.5$, $1$ e $3$ mm, and
in the case of compensation of the delay time between generated photons, as discussed later. Moreover we present an analysis
on the direct measurement of the visibility of the entangled state.
\par
\begin{figure}[!ht]
\centering
\includegraphics[width=0.44\textwidth]{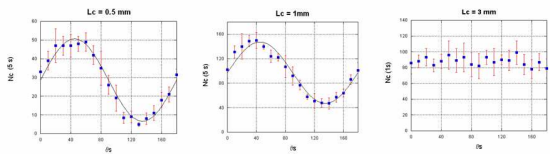}
\caption{Entangled state visibility as a function of the polarizer angle, for generating crystals of $0.5$, $1$ e $3$ mm thickness.} \label{f:corr}
\end{figure}

Data on correlation visibility are simply obtained by removing the HWP and QWP plates of the tomographic analyzer (see Fig.~\ref{f:exp}) and detecting the signal and idler
coincidence counts in a time interval, as a function of the signal
polarizer angle, and having fixed the idler polarization angle at $45^o$.
The theoretical prediction is:
\begin{align} P(\xi_s, 45^o_i) & = \:
_i\langle 45^o | \:\, _s\langle \xi_s | \: \rho \:
  | \xi_s \rangle_s \, | 45^o \rangle_i \, \nonumber \\ & = \,
  \frac{1}{2} p \: (\cos(\xi_s - 45^o))^2 + \frac{1}{4} (1-p)
   \label{correlazioni}
\end{align}
where $\xi_s$ is the angle of the signal polarizer in the
counter--clockwise direction, with the horizontal axis as the $0^o$
reference.
As it is apparent from this formula, when $p$ is near the unity (delay
time smaller with respect the coherence time of the pump) the
oscillating contribute due to the non--local correlations is dominant.
On the contrary, with greater delay time (and small $p$) the
correlations are washed out and the result is that of a statistical
mixture which does not depend on the angle. In particular, the maximum
of $P(\xi_s, 45^o_i)$ is at $45^o$, while the minimum is at $135^o$,
hence we can write explicitly the visibility $\mathcal{V}$ of the
oscillation as:
\begin{equation}
   \mathcal{V }= \frac{P(45^o_s, 45^o_i) - P(135^o_s, 45^o_i)}
   {P(45^o_s, 45^o_i) + P(135^o_s, 45^o_i)} = \, p
   \label{visibilita}
\end{equation}

\begin{figure}[!ht]
\centering
\includegraphics[width=0.44\textwidth]{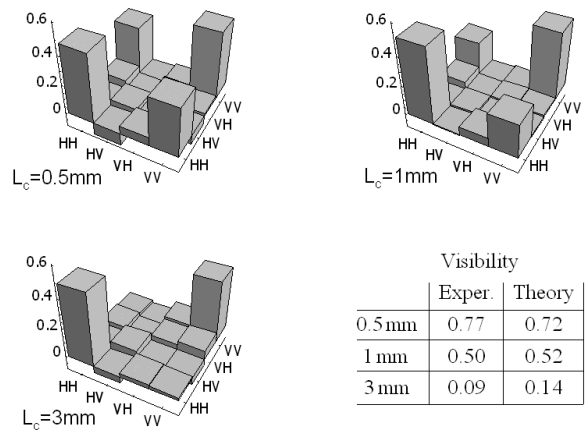}
\caption{Tomographic reconstruction of the generated
state for three different crystals. The measured and calculated
visibility are shown in table.} \label{f:rec}
\end{figure}

\begin{figure}[h!]
\centering
\includegraphics[width=0.44\textwidth]{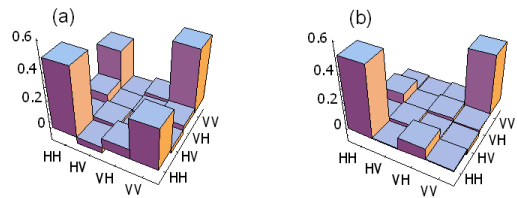}
\caption{Tomographic reconstruction with a delay compensation
crystal (see text). (a) Crystal angle set for maximum compensation,
visibility 0.66. (b) Crystal angle at $90^o$ with respect to (a),
visibility 0.17.} \label{f:comp}
\end{figure}

In Fig. \ref{f:corr} we show the visibility measurements as a function of the signal polarizer angle for the three different crystal pairs, with the theoretical prediction of eq.~(\ref{visibilita}) indicated by a full line.
The comparison between the theoretical density matrix elements of eq.~(\ref{matrix2}) and their tomographic reconstruction from experimental data is shown in Fig. \ref{f:rec}. It is confirmed that the off diagonal elements tend to reduce in magnitude for larger crystal thickness; in particular for $3$ mm crystals we obtain the density matrix of a statistical mixture.

In our model the lack of visibility of the entangled state is fully
ascribed to the decoherence effect due to the fluctuating phase difference between $H$ and $V$ parts of the SPDC, depending on the delay $\Delta\tau$. Having a very
small area of the fiber collimator, we have neglected any
decoherence of spatial origin, which introduces a phase variation
depending on the detector viewing angle. In order to verify this
statement, we have performed a series of measurements with the 3 mm
crystal, putting windows of 0.5 mm linear aperture in front of the
collimators: if the decoherence had a spatial contribution, we would
have expected an increasing in the state purity. In fact, the results of
the state reconstruction were the same as the original
configuration, thus supporting our hypothesis.

This fact also suggests how to improve the purity of the entangled
state by a phase retardation on the $H$ polarized part of the pump
with respect to the $V$ polarized part, to get $\delta_H(t +
\Delta\tau) = \delta_V (t)$. This can be approximatively
accomplished by inserting, along the pump ray and before the state
generator, a properly oriented BBO crystal with a suitable length.
We performed a series of measurements using the 1 mm double crystal
as state generator, and a 3 mm single crystal as pump phase
retarder. By varying the orientation of the axis of this crystal, we have
compensation or enhancement of the effect of the time delay between
the parts of the generated entangled state. In particular, the
visibility is expected to vary from a maximum to a minimum for a
$90^o$ change in orientation, as confirmed by the tomographic
reconstruction shown in Fig.~\ref{f:comp}. Notice that the maximum
visibility of 0.66 is larger than the corresponding visibility
without the auxiliary crystal (see Fig.~\ref{f:rec}), thus
demonstrating a partial time delay compensation.

\section{Measurements on the violation of Bell's inequality}
\label{s:bel}
We have also performed a series of measurements of the S parameter, characterizing the Bell's inequality in the CHSH version \cite{chsh}, for a comparison with the prediction of our theoretical model. To obtain reliable data on the S parameter we used the same experimental apparatus previously employed for correlation measurements. We considered as usual the 16 different configuration of the polarization angles on the signal an and on the idler \cite{deh02}.

\begin{figure}[h!]
\centering
\includegraphics[width=0.44\textwidth]{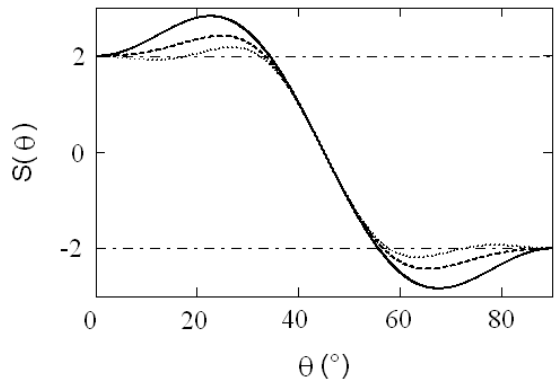}
\caption{S parameter as a function of $\theta$ for three different visibility values. Full line: $\mathcal{V} = 1$; dashed line; $\mathcal{V} = 0.7$; dotted line $\mathcal{V} = 0.5$.} \label{f:S-vs-theta}
\end{figure}

The Bell S parameter is theoretically defined as:
\begin{equation}
   S = E(a,b) - E(a,b') + E(a',b) + E(a',b')
   \label{parS}
\end{equation}
where the arguments $a, a'$ and $b, b'$ are the selected angles for signal polarizer and idler polarizer, respectively. The function $E$ is defined as $E(\alpha,\beta) = P(\alpha,\beta) + P(\alpha^{\perp},\beta^{\perp}) - P(\alpha,\beta^{\perp}) -
P(\alpha^{\perp},\beta)$, where $\alpha^{\perp} = \alpha + 90^o$ e
$\beta^{\perp} = \beta + 90^o$. The function $P$ is exactly that described in eq.~\ref{correlazioni}, but with the idler angle specified by the argument. For any realistic local theory one has $|S| \leq 2$, while for quantum mechanics $|S|$ can be greater than $2$, reaching a maximum value of $2 \sqrt{2}$. The following choice for the angles is used: $a = 0^o$, $b = \theta$, $a' = b + \theta$ and $b' = a' + \theta$. In this way $S$ is a function of the angle $\theta$ alone. In Fig. \ref{f:S-vs-theta} we show the calculated $S(\theta)$ for three states with different visibility $\mathcal{V} = p = (1, 0.7, 0.5)$; by decreasing $p$, the values of $S$ tend to
return in the limit of a local theory.

\begin{figure}[h!]
\centering
\includegraphics[width=0.44\textwidth]{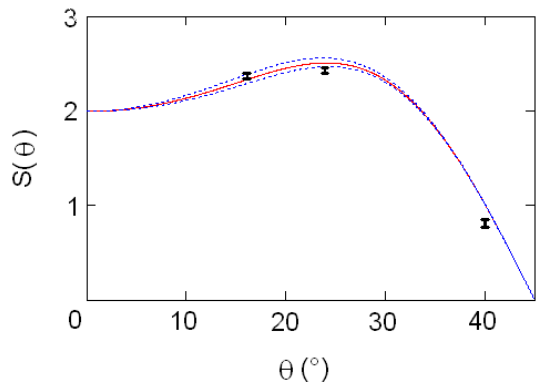}
\caption{Experimental results for S parameter for three values of $\theta$, compared with the theoretical S for a visibility of 0.77.} \label{f:Smisurato}
\end{figure}

For a comparison with these results of our model, we have measured the $S$ parameter for three different angles using as state generator the pair of $0.5$ mm crystals, that is the case with higher visibility. In Fig.~\ref{f:Smisurato} we show the theoretical curve of the $S$ parameter for a visibility equal to $0.77$ (full line), together with two other curves (dashed lines) indicating the extremal of the experimental errors, relative to the limited number of count during data acquisition. The three measurements of $S$ for the angles of $16^o, 24^o, 40^o$ are indicated with error bars. In particular we get $S(16^o) = 2.38 \pm 0.03$, $S(24^o) = 2.417 \pm 0.025$ and $S(40^o) = 0.80 \pm 0.05$. From these data we can conclude that in the case of $24^o$ the Bell's inequality is violated for more than $17$ standard deviations.

\section{Conclusions}
\label{s:out}
We have analyzed, both theoretically and experimentally, the generation
of polarization-entangled photon pairs by parametric down--conversion
from solid state CW lasers with small coherence time. In particular, we
have analyzed in some details a compact and low-cost setup  based on a
two-crystal scheme with Type-I phase matching. The effect of pump
coherence time on the entanglement and the nonlocality has been studied
as a function of the crystals length. The full density matrix has been
reconstructed by quantum tomography and the proposed theoretical model
is verified using a purification protocol based on a compensation
crystal.  We conclude that  laser diodes technology is of interest in
view of large scale application and that its that the characterization
of the generated state allows one to suitably tailor entanglement
distillation protocols.

\section*{Acknowledgements}
MGAP thanks Maria Bondani e Marco Genovese for useful discussions.
We are also indebted with Stefano La Torre for his help in detector
realization.

\appendix
\section{Measurement of the coherence length}
In this Appendix we experimentally verify that the spectrum of the down--converted signal is reduced when coincidence photon counts are performed, as a consequence of the trasverse momentum conservation.

If we observe only a single photon of the generated pair, the part of the spectrum incident on the coupling device is described in practice by the mismatch function $f(\omega_s)$ defined in Eqs. (\ref{ff}) and (\ref{fo}). But if we observe both photons and measure the simultaneous counts between signal and idler, we will detect a spectrum with a smaller width, and therefore we have a greater coherence length of the radiation. This because if we have a very wide spectrum for the signal at a fixed angle of observation, the idler photons, correlated with the signal photons also by transverse momentum conservation,  will be dispersed over an angle that can be wider with respect to the acceptance angle of the coupling device. Hence the pair of coupling devices work as a filters limiting the spectral window for observation. To the purpose of a determination of this effective spectral width, we present here some measurements using interference methods. In particular we performed two series of measurements, the first relative to the direct counts on a single detector to find the width associated with $f(\omega_s)$, the second relative to the coincidence counts on the two detectors to determine the width of the corrected mismatch function $\tilde{f}(\omega_s)$ used in eq.~(\ref{wf1}).

\begin{figure}[h!]
\centering
\includegraphics[width=0.44\textwidth]{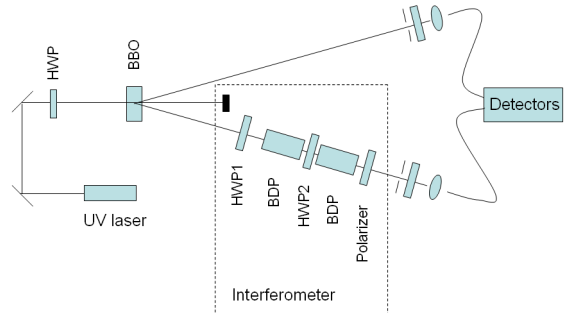}
\caption{Sketch of the experimental apparatus for the measure of the coherence length.} \label{f:schema-coerenza}
\end{figure}

In Fig. \ref{f:schema-coerenza} we show the experimental scheme (based on a single BBO crystal) employed for these
types of measurements. An interferometer equal to that described in \cite{gog05} is placed among the signal ray.
This interferometer is easy to align and is a very stable device. In both types of counting measurements we expect
to see interference fringes as a function of the delay time introduced by the interferometer between two optical
paths, and within the radiation coherence time. In particular we would determine a greater coherence length in the
case of coincidence counts with respect to the case of signal single counts.

The theoretical description of the interferometric experiment is as
follows. In the case of a single photon observation, and with a
crystal generating horizontal photons, the density matrix for the
signal before the interferometer can be built with the wavefunction of eq.~(\ref{propaga2}) of Section \ref{s:wav} by neglecting vertical polarization states, and tracing over the idler frequency:

\begin{equation}
  \rho_1 \, = \, \int \, d\Omega \,
   |f(\Omega^0 + \Omega)|^2 \,\,  |H, \Omega \rangle_s \,\, _s \langle
   H, \Omega |
\end{equation}
where we do not consider the immaterial propagation factor and use the original mismatch function of eq.~(\ref{fo}). 

The density matrix after the interferometer follows by considering:
(a) a polarization rotation of $45^o$ due to the HWP plate placed
before the first calcite crystal; (b) the delay time $\tau$
introduced by the interferometer between the $H$ and $V$ parts; (c)
the projection of these states over the axis of the final polarizer
oriented at $45^o$, placed before the coupling device. The final
density matrix is easy obtained as:

\begin{equation}
  \rho_1 \, = \, \int \, d\Omega \,
   |f(\Omega^0 + \Omega)|^2 \, \frac{1}{4} \, 
   \left| 1 + e^{i \Omega \tau} \right|^2 \,  
   |45^o, \Omega \rangle_s \,\, _s \langle
   45^o , \Omega |
\end{equation}
The probability to observe a count on the detector is then proportional to:

\begin{align}
   P_1(\tau) \, i& = \, \int \, d\Omega^{'} \, _s\langle 45^o,
   \Omega^{'}| \rho_1 | 45^o, \Omega^{'} \rangle_s
 \nonumber \\  & = \,
\int \, d\Omega \,
   |f(\Omega^0 + \Omega)|^2 \, \frac{1}{4} \, \left| 1 + e^{i \Omega \tau} \right|^2
   \label{prob1}
\end{align}
The width of the interference pattern representing count numbers as a function of the delay $\tau$, is given by a factor similar to a Fourier transform of the down-converted power spectrum; hence this width scales as the inverse of the spectral power width of the function $f$.

In the case of signal and idler coincidence counting, the state vector is again derived from eq.~(\ref{propaga2}), by taking only the $H$ part and discarding the propagation factor. After the passage in the interferometer, it is straightforward to see that the coincidence probability is the same as for the single count probability by replacing $f(\Omega^0 + \Omega)$ with the modified mismatch function $\tilde{f}(\Omega^0 + \Omega)$. 
Hence in this case the width of the interference curve is governed by the modified spectral power width $\Delta \omega_s$ of $\tilde{f}$.

\begin{figure}[h!]
\centering
\includegraphics[width=0.44\textwidth]{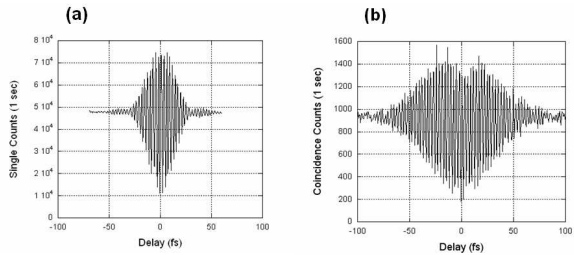}
\caption{Interference patterns: (left) single signal counts; (right) signal and idler coincidence counts.} \label{f:confronto-corr}
\end{figure}

In Fig. \ref{f:confronto-corr} we show on the left the interference pattern obtained with signal single counts, using the BBO crystal of 3 mm length. The width of the curve is near $30$ fs, corresponding to a down converted spectrum of about $64$ nm. On the right we show the pattern in the case of signal and idler coincidence counts: the coherence time is enlarged to $70$ fs, corresponding to a spectral width of $27$ nm. In both cases the coherence length is well below that of the pump light. These data are used to determine the appropriate correction factor $R(\Delta \omega_s) = \tilde{f}(\omega_s) / f(\omega_s)$ in the definition of the wavefunction eq.~(\ref{wf1}).

\section{Complete calculation of the delay times in state generation }
In deriving the delay time between $(HH)$ and $(VV)$ photons, we assumed state generation in the crystals middle. But in fact these states can be generated in any point in the inner of the crystals, therefore the propagation factors $P(H)$ and $P(V)$ must be position dependent. Let's indicate with $z_1$ and $z_2$ the longitudinal coordinates of the internal generation points for the first crystal and for the second crystal, respectively. Referring to the Fig.~\ref{f:delay}, we now have the following two equations for the propagation times $\tau_H$ and $\tau_V$:
\begin{equation}
    \tau_H(z_1,z_2) = \frac{L_C - z_1}{V_p^o} + \frac{z_2}{V_p^e} +
    \frac{L_C - z_2}{V^o \cos(\phi_3)}
\end{equation}

\begin{equation}
    \tau_V(z_1,z_2) = \frac{L_C - z_1}{V^o \cos(\phi_1)} + \frac{L_C}{V^e \cos(\phi_2)}
\end{equation}

Generally speaking, the state visibility $p$ would depend on the delay time $\Delta \tau(z_1,z_2) = \tau_H(z_1,z_2) - \tau_V(z_1,z_2)$. Because we do not have any information about the effective position in which a particular photon pair is generated, we consider an average over the possible positions, by integrating with a flat distribution probability:
\begin{equation}
    p_z = \frac{1}{{L_C}^2} \, \int dz_1 \, dz_2 \,\,
    e^{- \Delta \tau(z_1,z_2) / \tau_c}
\end{equation}

\begin{figure}[h!]
\centering
\includegraphics[width=0.44\textwidth]{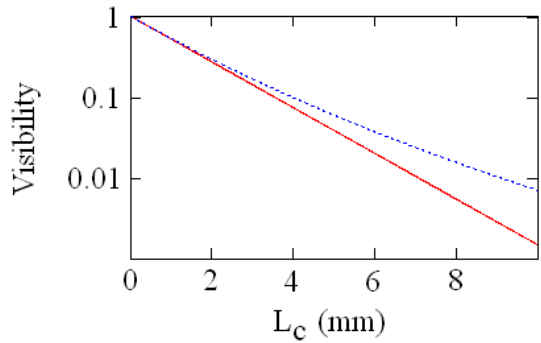}
\caption{Entangled state visibility as a function of the crystal length for the different assumptions on the state generation position. Full line: in the crystals middle; Dotted line: in the whole crystal length.} \label{f:delay-z}
\end{figure}

In Fig.~\ref{f:delay-z} we show a comparison between the visibility for state generation in the crystals middle, and that obtained from the above formula. It is clear that there is some difference only for very small visibilities obtained with very long crystals.

\end{document}